# Molecular Magnetocapacitance

Yu-Ning Wu[1], X.-G. Zhang[2], and Hai-Ping Cheng[1*]

[1]Department of Physics and the Quantum Theory Project, University of Florida, Gainesville, FL 32611

[2] Center for Nanophase Materials Sciences, Oak Ridge National Laboratory, Oak Ridge, TN 37831-6493

**Capacitance of a nanoscale system is usually thought of having two contributions, a classical electrostatic contribution and a quantum contribution dependent on the density of states and/or molecular orbitals close to the Fermi energy. In this letter we demonstrate that in molecular nano-magnets and other magnetic nanoscale systems, the quantum part of the capacitance becomes spin-dependent, and is tunable by an external magnetic field. This molecular magnetocapacitance can be realized using single molecule nano-magnets and/or other nano-structures that have antiferromagnetic ground states. As a proof of principle, first-principles calculation of the nano-magnet [$Mn_3O(sao)_3(O_2CMe)(H_2O)(py)_3$] shows that the charging energy of the high-spin state is 260 meV lower than that of the low-spin state, yielding a 6% difference in capacitance. A magnetic field of ~40T can switch the spin state, thus changing the molecular capacitance. A smaller switching field may be achieved using nanostructures whose physical properties such as magnetic moment are size-dependent. Molecular magnetocapacitance may lead to revolutionary device designs, e.g., by exploiting the Coulomb blockade magnetoresistance whereby a small change in capacitance can lead to a huge change in resistance.**

*cheng@qtp.ufl.edu

Quantum mechanical effects can change the capacitance of a mesoscopic capacitor by a contribution due to the density of states [1], and one quantum consequence is magnetocapacitance due to the asymmetry in the capacitance tensor elements under field reversal [2]. On the nanoscale, quantum capacitance of a molecule may depend on the charge density distributions of the highest occupied molecular orbital (HOMO) and the lowest unoccupied molecular orbital (LUMO). For single molecule magnets (SMM) and other magnetic nanosystems whose HOMO and LUMO are determined by their magnetic states, it is natural to expect that their self-capacitances are in turn dependent on their magnetic states. Such an effect, if proven to exist, will provide a much simpler path to achieve magnetocapacitance in nanoscale materials. By merely switching the magnetic state of a molecular nano-magnet, one can change its capacitance.

In this letter, we demonstrate the concept of molecular magnetocapacitance based on first-principles calculations of single molecular nano-magnets. Molecular nano-magnets are stable at room temperature and can be crystallized or used in single-molecule tunneling junctions [3-7]. A rich array of magnetic states or spin states has been probed. Tunneling transport through a $Mn_{12}$ single-electron transistor has been studied using density functional theory recently. [8] Our model system is a single molecular nano-magnet [$Mn_3O(sao)_3(O_2CMe)(H_2O)(py)_3$], that contains three $Mn^{III}$ ions, the key for its magnetic properties, three pyridine ligands, one carboxylate group and a water molecule. For simplicity, we abbreviate the molecular formula as [$Mn_3$]. In experiments, this SMM can be in an $S=6$ high-spin (HS) state or in an $S=2$ low-spin (LS) state depending on the relative spin orientations of the three Mn$^{III}$ ions. The LS state is observed as the ground state in experiment. The system can also be viewed as a zero-dimensional quantum dot. Following the classical definition, we relate potential energy change upon charging and discharging to the capacitance of the system, but with a full quantum description of electrons coupled with molecular configurations. For nano-dots, the capacitance can be obtained by the ratio of the charge variation to the chemical potential variation. The important quantity is the charging energy $E_c$ (sometimes also called capacitive energy), which is the difference between the ionization potential (IP) and the electron affinity (EA),[9,10]

$$E_c = \frac{e^2}{C(N)} = IP(N) - EA(N), \quad (1)$$

where *N* is the number of electrons in the system, *IP* and *EA* are *the least energy* needed to subtract an electron from, and *the most energy* released to attach an electron to a system of *N* electrons, respectively. With this important *least-most energy principle* in mind, we examine carefully physical properties of [Mn$_3$]. The basic procedure consists of four steps as, 1) optimize molecular configuration and obtain electronic structure and magnetic pattern, 2) add and/or subtract an electron of spin-up and/or spin-down followed by optimization again, 3) extract $E_c$ according *the least-most energy* principle, and 4) calculate magnetic quantum conductance. Before proceeding to step 1, we define the magnetocapacitance as

$$MC = (C_{HS} - C_{LS})/C_{LS}, \quad (2)$$

following the definition of magnetoresistance. Here $C_{HS}$ and $C_{LS}$ are the capacitances of HS and LS states respectively.

We performed Kohn-Sham density functional theory (DFT)[11] calculations to investigate the ionization potential and electron affinity of the [Mn$_3$] SMM system. We used the spin-polarized Perdew-Burke-Ernzerhof (PBE) exchange-correlation functional in the PAW[12,13] pseudopotential formalism, which is implemented in the plane-wave based VASP[14,15] package. The [Mn$_3$] molecule was placed in a 35 Å by 35 Å by 35 Å unit cell for isolation from neighboring molecules for both of neutral and charged systems; thus only the Γ-point was used for first Brillouin zone[16]. The plane-wave energy cutoff was 500 eV. Thresholds for self-consistent calculation and structure optimization are set as $10^{-5}$ eV and 0.02 eV/Å, respectively. The polarizability is calculated by linear interpolation of the induced dipole moment and applied electric field. The dipole moment is calculated with the same criteria as self-consistent calculation with dipole correction.

Configuration optimization is essential to nano-systems that contain $10^2$-$10^3$ atoms because of the strong interplay between structure and properties. All calculations should be performed using same theoretical treatment for maximal error cancellations. Figure 1 shows the optimized structure of a [Mn$_3$] molecule. It can be seen that three pyridine ligands are attached to Mn$^{III}$ ions above the [Mn$^{III}$]$_3$-plane. Below the [Mn$^{III}$]$_3$ plane, one carboxylate group is shared by Mn2 and Mn3, while a water molecule is attached to Mn1. The rest of the atoms of the molecule lie almost in the [Mn$^{III}$]$_3$-plane. The largest deviation from the plane is the position of the middle oxygen (O1) atom[17,18] (0.39 Å above the plane, in good agreement with the experimental value of 0.33 Å), followed the position of one side oxygen (O2) atom (0.33 Å below the plane). The deviations of all other atoms range from 0 to 0.20 Å. Both the HS and LS states show very similar structures after optimization.

There are three distinct low-spin configurations, 1=(down, up, up), 2=(up, down, up) and 3=(up, up, down), but the second and the third ones are equivalent due to the system symmetry. The second LS configuration, or LS2, is predicted to be more stable than the first LS type by 21 meV. Our calculations also show that LS2 is energetically more stable than HS state by 37 meV, in good agreement with previous calculations.

Next we proceed in step 2 to examine energy changes in various initial and final states of [Mn$_3$] upon adding or removing an electron. We consider only the most stable LS state. Anions (cations) were prepared by adding (removing) a spin-up or a spin-down electron such that we created ions of all possible different spin states. Table 1 shows the energies of the neutral molecule, cation and anion of both HS and LS states. We denote anion_up and anion_down as gaining a spin-up and a spin-down electron, respectively. Similarly, cation_up and cation_down refer to losing a spin-up and spin-down electron, respectively. Structural optimizations were performed for all states. The relaxation energies, defined as the energy difference before and after structural relaxation for a charged system from the neutral structure are 33, 57, 78 and 54 meV for anion_up, anion_down, cation_up and cation_down in the HS state, while those of LS states are 88, 86, 79 and 73 meV. As shown, the HS state prefers to adsorb a spin-up electron over a spin-down electron by 75 meV, and favors losing a spin-up electron rather than a spin-down one by 592 meV. In contrast to the HS state, the LS state, which is the ground state, prefers to gain a spin-down electron over a spin-up electron by 45 meV. However, it prefers to give away a spin-up electron than a spin-down one by 80 meV.

Step 3 was to follow *the least-most energy* principle and select the most stable anion and cation states for calculations of ionization potential and electron affinity. Table 2 lists the Ionization potential ($IP$), electron affinity ($EA$), capacitance ($C$) and charging energy ($E_c$) of both HS and LS states (Step 4 followed immediately once the right $IP$ and $EA$ were identified). Charging energy and capacitance were calculated according to equation (1). It can been seen that the HS state is 175 meV lower in $IP$ and 75 meV higher in $EA$ than the LS state, resulting in a capacitance of the HS state that is 6% (or $0.247 \times 10^{-20}$ F) higher than in the LS state (or 6% magnetocapcitance), and a charging energy that is 260 meV lower than those of the LS state.

The difference in $E_c$ between high-spin and low-spin state constitutes the physical foundation for the concept of quantum magnetocapacitance. Without a magnetic field, the molecule stays in the LS ground state, which has a high charging energy. The system can be switched into the HS state, which has a lower charging energy, by applying a sufficiently high magnetic field, resulting in a change in the quantum capacitance of the molecule or a quantum magnetocapacitance. We estimate the magnitude of the switching magnetic field via $B = \Delta E / g\mu_B \Delta M$, where $\Delta E$ is the energy difference between LS and HS states (37 meV), $\Delta M$ is the magnetic moment difference between HS and LS states, $\mu_B$ =0.058 meV/T, and the *g*-factor is equal to 2. With these values, the switching magnetic field is approximately 40 T at 0 K.

It is important to understand the microscopic origin of the charging energy difference between HS and LS states. We thus calculated the spatial distribution of total charge difference between the neutral and the charged [Mn$_3$] for both anions and cations in the HS and LS states. Figure 2, panels (a) and (b) depict the difference between the neutral molecule and the anion in the HS state, and panel (c) and (d) show those in the LS state. By comparing Fig. 2 with electron orbitals (not shown, see supplementary materials Fig. A), we fond that the charge density difference is mainly from the highest occupied electron orbitals (HOMOs). Note that the electron in the HOMO of the neutral molecule is the electron lost in the ionization process and the electron in the HOMO of the anion is the electron gained when attaching an electron. Panels (a) and (c) (correspond to HOMOs of the HS and LS neutral atoms, respectively) show significant difference between the HS and the LS cations, especially at the Mn2 site. Drastically different distributions of the lost electron between the HS and the LS states lead to a relatively large difference in *IP* (175 meV). Meanwhile, panels (b) and (d) (correspond to the HOMOs of the HS and the LS anions, respectively) display some similarities, especially on all three Mn atoms, which explains the relatively small difference in *EA* (75 meV). The main difference is that in panel (b), the center oxygen atom has more charge than the one in panel (d).

The mechanism of quantum magnetocapacitance is therefore clear: the charging process in a magnetic system depends on the magnetic state of the system and also on the spin of the incoming and out-going electrons. The capacitance can be controlled by external magnetic field, by changing the spin configuration of a quantum dot. We stress here that the proposed controllable magnetic quantum capacitance is fundamentally different from tuning the quantum capacitance by utilizing Landau levels [19]. There, the system itself is non-magnetic and thus the capacitance is not spin dependent. As the size of a system is reduced, it becomes harder and harder to utilize Landau levels. To generate one magnetic flux quantum through a quantum dot of 2x2 nm$^2$ (the size of our molecule) in cross-sectional area, such as the one in our study (in the *x-y* plane), the required magnetic field is 500 Tesla. The switching field for our model molecule of about 40 T does not allow even one electron in each Landau level, and the capacitance cannot be modulated through Landau levels under such a field.

This type of molecular magnetocapacitance is best exploited through the Coulomb blockade effect. Recently it has been proposed [20,21] that a small spin-dependence of the charging energy of a quantum dot can lead to a giant Coulomb blockade magnetoresistance effect. Molecular magnets and magnetic nanostructures that demonstrate magnetocapacitance are the perfect candidates for realizing this effect.

The concept of the quantum self-capacitance should be distinguished from the polarizability of a molecule, even though both are related to the concept of a capacitance at some level. The polarizability is only a factor affecting the mutual capacitance between the source and the drain if such molecules are used as a dielectric medium. The charging energy, on the other hand, is the essential quantity in the Coulomb blockade effect whereby an electron is injected onto the molecule. The pertinent capacitance in the latter case is the self-capacitance of the molecule (or more precisely the mutual capacitance between the molecule and an electrode). We have shown that the quantum self-capacitance has a strong spin-dependence. It is natural to further ask whether the molecular polarizability also has a similar spin-dependence. We have performed dielectric constant and polarizability calculations of the molecule in the HS and the LS states within the linear response regime, that is, we assume a linear dependence between the dipole moment and applied electric field. By comparing the calculated response tensor elements, we have found that the maximum difference between the LS and HS states is less than 0.5%. The sharp contrast between the energy calculations and the polarizibility (or dielectric constant) calculations highlights the different physics represented by these two quantities. The polarizability reflects how all electrons collectively respond to an external field, whereas the quantum capacitance is mainly determined by only the HOMO and LUMO orbitals. Therefore, a molecule may have different self-capacitances in two spin states but a spin-independent polarizability. It is clear then that when the molecule is used as a dielectric medium its magnetic moment does not affect the polarizability. When it is used as a quantum dot for Coulomb blockade a strong spin-dependence in the current should appear. The capacitances for these two applications are entirely different.

Finally, the proposed magnetic quantum capacitor can be realized by nano-structure other than SMM; for example, a system that consists of two Fe particles separated by a $C_{60}$ molecule (see supplementary materials Fig. B) can have an AFM ground state. Our calculations show that the energy difference between the AFM state and the FM state is a function of $Fe_n$ particle size. Beyond certain size, the ground state of the system transits from the AFM state to the FM state (see supplementary materials Table A and B). The estimated switching field can be substantially reduced (see supplementary materials Table C) by increasing the size of the attached Fe clusters (or the magnetic moments).

In summary, we have demonstrated the concept of molecular magnetocapacitance that the capacitance of nano-magnet can be spin-dependent. As an example, the Mn3 molecular nano-magent has been investigated by first-principles calculations. The magnetocapacitance of Mn3 molecule is determined as 6%, and a 40T magnetic field is needed to switch the molecule from low-spin state to high-spin state. The proposed $Fe_n$-$C_{60}$-$Fe_n$ systems are also good candidates for molecular magnetocapacitance, in which the switching field can be lowered by increasing the magnetic moment. Our findings on SMM and $Fe_n$-$C_{60}$-$Fe_n$ suggest that a relentless search for candidate systems can be very fruitful. Future synthesis of SMM guided by the energy principle may hold the key for realizing quantum dots with capacitance that is tunable using magnetic field under 1 T.

Acknowledgement: This work is supported by U. S. DOE/BES-FG02-02ER45995. A portion of this research was conducted at CNMS sponsored at ORNL by the Scientific User Facilities Division, Office of Basic Energy Sciences, U.S. DOE. The authors also acknowledge NERSC and UF-HPC for computing resources.

Figure captions:

Fig.1: Optimized structure of [Mn$_3$] molecule. Panel (a) is top view and panel (b) is side view. Mn atoms are in purple. O atoms are in red. Blue spheres are N atoms. Grey and blue-green sticks stand for C and H atoms.

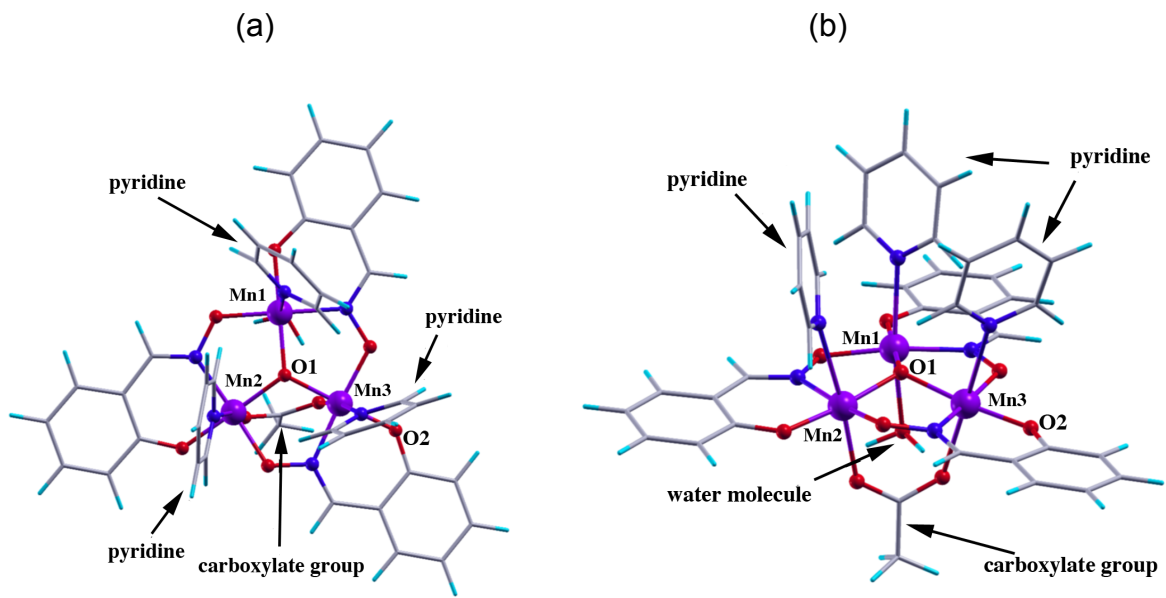

Fig. 2: Isosurfaces of charge difference of (a) between neutral molecule and cation of high-spin state, (b) between anion and neutral molecule of high-spin state, (c) between neutral molecule and cation of low-spin state and (d) between anion and neutral molecule of low-spin state. Isovalue is 0.015 e/Å$^3$.

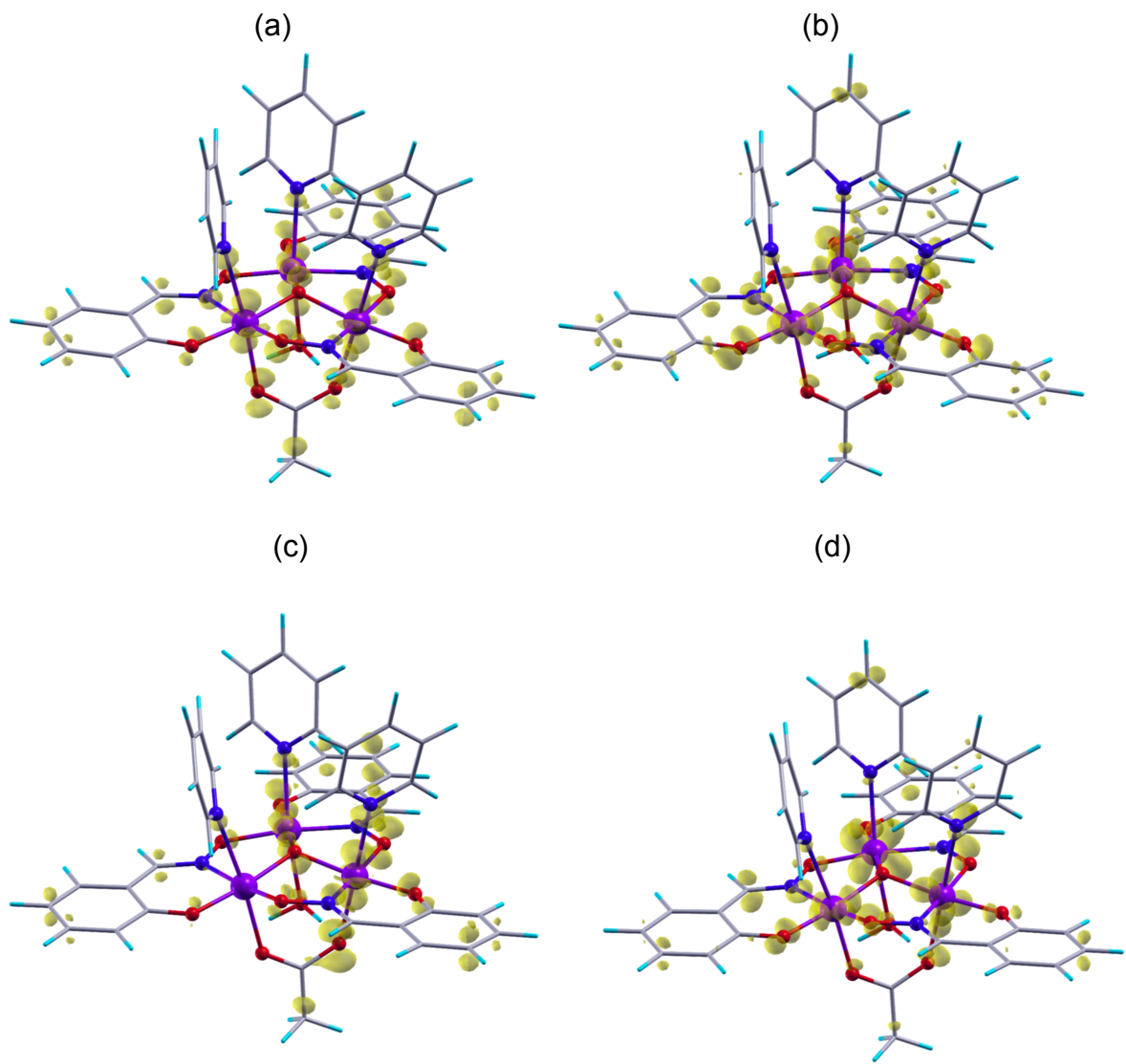

Table 1: Energies of neutral case, cation and anion of both HS and LS states. Adding/removing one spin up/down electron are all considered. The energy of neutral LS state (ground state) is set to be 0.

|  | High-spin state | | Low-spin state | |
| --- | --- | --- | --- | --- |
|  | Energy (eV) | Magnetization ($\mu_B$) | Energy (eV) | Magnetization ($\mu_B$) |
| neutral | 0.037 | 12 | 0 | 4 |
| anion_up | **-1.627** | 13 | -1.544 | 5 |
| anion_down | -1.552 | 11 | **-1.589** | 3 |
| cation_up | **5.677** | 11 | **5.825** | 3 |
| cation_down | 6.269 | 13 | 5.905 | 5 |

Table 2: Ionic potential (IP), electron affinity (EA), capacitance (C) and charging energy ($E_c$) of both HS and LS states.

|  | High-spin state | Low-spin state | $2(HS-LS)/(HS+LS)$ |
| --- | --- | --- | --- |
| IP (eV) | 5.640 | 5.825 | -3% |
| EA (eV) | 1.664 | 1.589 | 4% |
| C ($10^{-20}$ F) | 4.029 | 3.782 | 6% |
| $E_c$ (eV) | 3.976 | 4.236 | -6% |

Supplementary information

Fig.A: Isosurfaces of charge density of (a) high-spin neutral state HOMO, (b) high-spin anion state HOMO, (c) low-spin neutral state HOMO and (d) low-spin anion state HOMO. Isovalue is 0.015 e/Å$^3$.

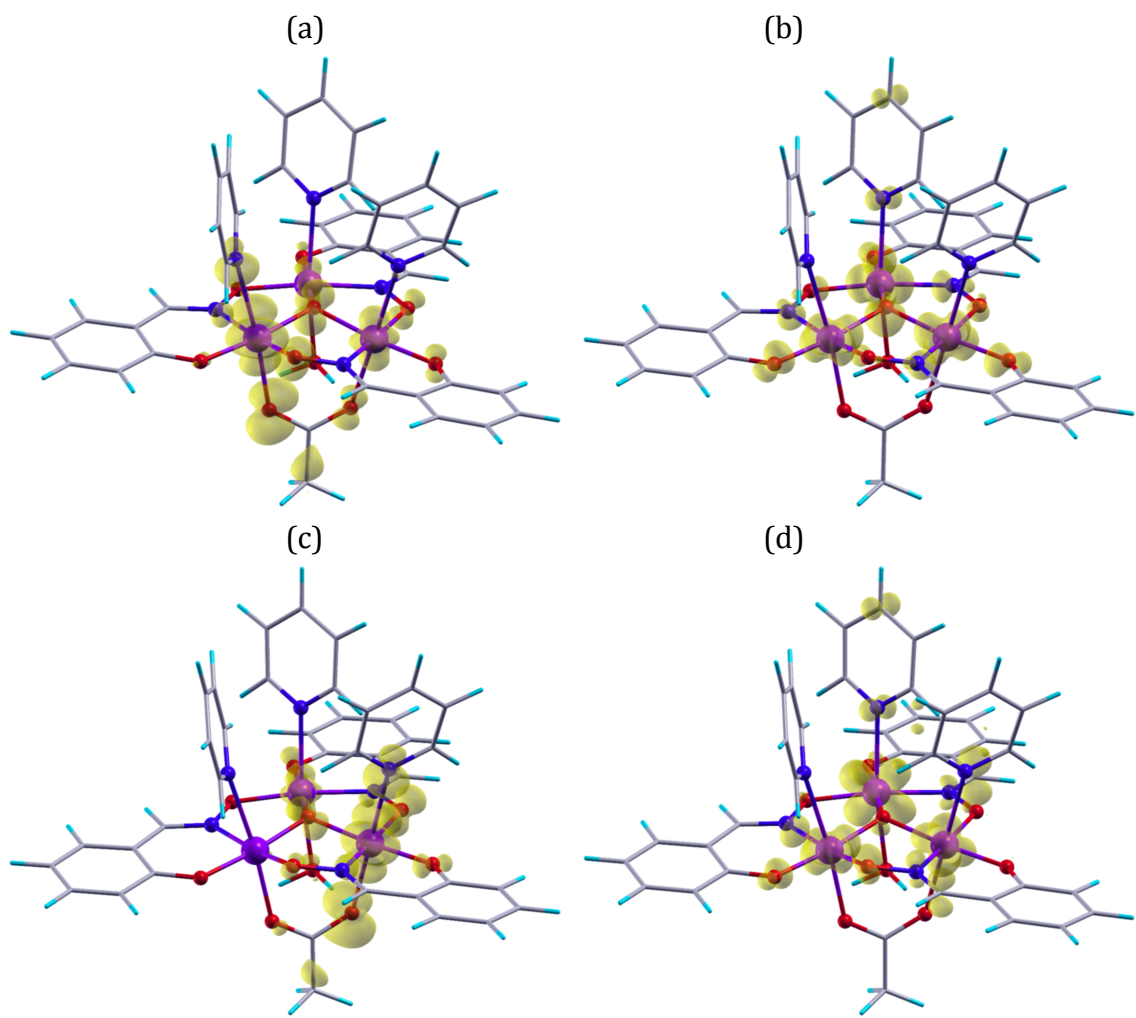

Fe$_n$-C$_{60}$-Fe$_n$ system is another candidate for the quantum magnetocapacitance. The structure optimization and energy calculations are performed with the same criteria as Mn$_3$ molecule. Table A and B present the energetic and capacitance information of Fe-C$_{60}$-Fe, corresponding to Table 1 and 2 for Mn$_3$ molecule in the main text. AFM state shows 321 meV lower than FM state in charging energy, indicating a 5.5 percent higher capacitance for AFM state. As shown in Table C, the estimated switching field is 124.4 T for Fe-C$_{60}$-Fe system. By enlarging the attached Fe clusters, the magnetic moment difference between AFM and FM states increases and the energy difference decreases. This leads to the drop of the switching field. The switching field of Fe$_{15}$-C$_{60}$-Fe$_{15}$ is estimated as 1.2 T. Our calculation also indicates the FM state will be the ground state for Fe$_{50}$-C$_{60}$-Fe$_{50}$, so the size of the attached Fe clusters cannot beyond certain size for magnetocapacitance.

Fig.B    C$_{60}$ with 2 Fe atoms attached. C atoms are in yellow and Fe atoms are in red.

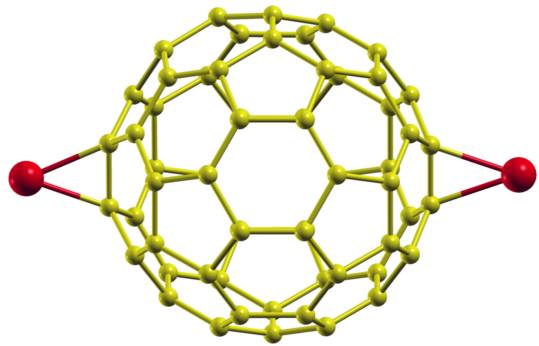

Table A: Energies of neutral case, cation and anion of both AFM and FM states. Adding/removing one spin up/down electron are all considered. The energy of neutral AFM state (ground state) is set to be 0.

|  | AFM | | FM | |
| --- | --- | --- | --- | --- |
|  | Energy (eV) | Magnetization ($\mu_B$) | Energy (eV) | Magnetization ($\mu_B$) |
| neutral | 0.0 | -0.269 | 0.083 | 6.022 |
| anion | -2.558 | 1.033 | -2.641 | 6.573 |
| cation | 5.402 | 0.023 | 5.498 | 5.300 |

Table B: The ionization potential, electron affinity, charging energy and capacitance of the Fe-$C_{60}$-Fe system for both AFM and FM states.

|  | AFM | FM | $2(FM - AFM)/(FM + AFM)$ |
| --- | --- | --- | --- |
| IP (eV) | 5.402 | 5.415 | 1% |
| EA (eV) | 2.558 | 2.724 | 13% |
| C ($10^{-20}$ F) | 5.633 | 5.954 | 6% |
| $E_c$ (eV) | 2.844 | 2.691 | -6% |

Table C: Energies, magnetizations and estimated switch fields (from AFM to FM) of $Fe_n$-$C_{60}$-$Fe_n$ systems. The switching field decreases as the number of Fe atoms (or the magnetic moment) increases. $Fe_{50}$-$C_{60}$-$Fe_{50}$ turns out to have a FM ground state.

|  | AFM | | FM | | Switching Field (T) |
| --- | --- | --- | --- | --- | --- |
|  | Energy (eV) | Magnetization ($\mu_B$) | Energy (eV) | Magnetization ($\mu_B$) |  |
| Fe-$C_{60}$-Fe | 0.0 | -0.27 | 0.083 | 6.02 | 124.4 |
| $Fe_{10}$-$C_{60}$-$Fe_{10}$ | 0.0 | 0.00 | 0.022 | 53.23 | 3.6 |
| $Fe_{15}$-$C_{60}$-$Fe_{15}$ | 0.0 | 0.00 | 0.012 | 86.73 | 1.2 |
| $Fe_{50}$-$C_{60}$-$Fe_{50}$ | 0.014 | 0.91 | 0.0 | 281.34 | N/A |